# GreyShot: Zeroshot and Privacy-preserving Recommender System by GM(1,1) Model

*Hao Wang*

*CEO Office, Ratidar Technologies LLC, Beijing, China，100011*

A B S T R A C T

Every recommendation engineer needs to face the cold start problem when building his system. During the past decades, most scientists adopted transfer learning and meta learning to solve the problem. Although notable exceptions such as ZeroMat etc. have been invented in recent years, cold-start problem remains a challenging problem for many researchers. In this paper, we build a zeroshot and privacy-preserving recommender system algorithm GreyShot using GM(1,1) model by taking advantage of the Poisson-Pareto property of the online rating data. Our approach relies on no input data and is effective in generating both accurate and fair results. In conclusion, zeroshot problem of recommender systems could be effectively solved by grey system methods such as GM(1,1).

## 1. Introduction

Recommender system is generating a large amount of traffic volume on the internet nowadays. Companies such as Amazon and Toutiao have more than 30% of their traffic volume attributed to the recommendation technology. Investment in recommender system has been skyrocketing because people are building larger and deeper models to solve real world problems. Since 2016, when deep neural network models started to dominate major research venues such as ACM RecSys, deep learning has become the de facto standard for the industry.

A commonly cited example to demonstrate the basic idea of recommendation technology is that the algorithm use similar users' view log to recommend new items to a user. This idea is actually the earliest recommendation algorithm named collaborative filtering. The algorithm was invented in 1992 by D. Goldberg et.al. [1], but was proved wrong in 2024 by H. Wang [2]. The technology dominated the academia and industry in the first decade of the technology's history and it is unfortunate that 32 years after its debut it was debunked.

A challenging problem faced by every recommender system scientist is the zeroshot problem, namely, when a new item or a new user enters the system, how could we recommend items to the user ? There is no historic data that we could rely on, so many researchers resort to transfer learning/meta learning to borrow insight from other domains.

In 2021, a new algorithm named ZeroMat [3] was invented, and a series of algorithms such as RankMat [4], PowerMat [5], PoissonMat [6] and LogitMat [7] followed suit in the next couple of years. This paradigm of algorithms fully utilize the power law effect of the data structures, and for the first time in human history, recommend items without actual reference to real user item rating data or data/models from other domains.

GM(1,1) model [8] is a grey system model that predicts time series data. The model is especially suited for time series whose individual values need not to be super nice for the task but the partial sums of the database exhibit exponential behavior. GM(1,1) and its variants [9-12] have been widely applied in a series of application scenarios including recommender system [13] - the topic of this paper.

By inspecting on the Poisson-Pareto behavior of the online rating data, which could be modeled by individual votes generated by non-homogeneous Poisson process, we believe that GM(1,1) is a proper model to model the user rating data. Therefore, we could apply GM(1,1) in our recommender system problems.

In this paper, we introduce a new paradigm of algorithms that uses no transfer learning or meta learning approaches but solves the Experiment zeroshot learning problem using GM(1,1) model. Our robust and privacy-preserving algorithm is named GreyShot. We prove by experiments that our algorithm is competitive with other zeroshot algorithms by both accuracy metric and fairness metric.

* *Corresponding author:* Hao Wang
  E-mail address: Haow85@live.com

## 2. Related work

Recommender system was invented in 1992 by D.Goldberg et.al. [1] The first recommendation algorithm was named Collaborative Filtering. Although it has been the most famous algorithm in the field, it was proved to be wrong by H.Wang [2] in 2024. The next wave of technical revolution in the field was initiated by the Netflix Prize Contest. Notable inventions in this period include probabilistic matrix factorization [14], SVDFeature [15], SVD++ [16], and many other variants based on matrix factorization.

Although linear models [17][18] had been popular in the industry due to its complexity and speed, deep neural networks [19][20][21] started to flood the field in 2016. DLRM [22], DCN [23] and DeepFM [24], among a whole spectrum of techniques, stood out as the de facto standards of the industry.

Data analysis of recommender systems is also an important subfield of the research area. H.Wang [25] proved that online rating behavior is essentially a non-homogeneous Poisson process. He [26] further points out that the parameter space of matrix factorization is not Gaussian.

In recent years, research on recommendation fairness has aroused public awareness. A.Beutel [27] proposed Focused Learning to solve the popularity bias problem. Similar regularization techniques have been widely adopted in the field [28][29][30].

Grey System Modeling is a field invented by Chinese scientists [31]. The theory has been applied in environment protection [32], energy industry [33] and education [34], etc. [35][36][37], its application in the field of recommender system is new [13][38]. In this paper, we use Grey System Modeling combined with matrix factorization technique to solve the cold-start and privacy problem of recommender systems.

## 3. System modeling

We now apply grey system modeling to solve the recommendation problem and name our new algorithm GreyShot. We model the user item rating behavior as a collective value of individual rating behavior as follows: The user item rating value of user i on item j is $R_{i,j}$. We treat the value of $R_{i,j}$ as votes received by users. For example, Mr. Smith gave *Black Myth: Wu Kong* a 5 star on a game rating platform. This is treated as 5 events were happening - each 1 star represents an event. Researchers have modeled such events as Poisson process. By decomposing each star value into separate voting behavior, we are able to model the likelihood of receiving such a star value under the Poisson assumption.

We treat the voting process for each star value as a time series array, and the star value is the partial sum over time of the time series values. Formally, we define the voting process as :

$$r_u^{(0)} = \{r^{(0)}(t)\}$$

, where $r^{(0)}(t)$ is distributed in line with a non-homogeneous Poisson process and takes only the value of 1. The user item rating value, i.e., the partial sums of $r_u^{(0)}$ is defined as follows :

$$r_u^{(1)} = \{r^{(1)}(t)\}$$

, where :

$$r^{(1)}(t) = \sum_{i=1}^{t} r^{(0)}(i)$$

We now resort to the GM(1,1) model to model the partial sum series :

$$\frac{dr^{(1)}(t)}{dt} + aZ^{(1)} = b$$

, where $Z^{(1)}(t) = \alpha r^{(1)}(t) + (1-\alpha)r^{(1)}(t+1)$. Solving for $r^{(1)}(t)$ by GM(1,1), we have:

$$R_{i,j} = r^{(1)}(t+1) = \left(r^{(0)}(1) - \frac{b}{a}\right)e^{-at} + \frac{b}{a}$$

Since we assume the voting events is a Poisson process, the number of events happening within a time interval is proportional to the length of the interval. Based on this and GM(1,1) model. Also $r^{(0)}(i)$ takes only the value of 1, we obtain the following relation:

$$R_{i,j} = \left(1 - \frac{b}{a}\right)e^{-aR_{i,j}} + \frac{b}{a} \qquad (1)$$

In the classic matrix factorization paradigm, the computation of user item rating value is defined as follows [8]:

$$R_{i,j} = U_i^T \cdot V_j \qquad (2)$$

By Power Law Effect, we have the following total probability distribution of the dataset:

$$L = \prod_{i=1}^{N}\prod_{j=1}^{M} R_{i,j}^{R_{i,j}} \qquad (3)$$

Plug Equation (1) into Equation (3), we obtain the loss function for the GreyShot algorithm:

$$L = \prod_{i=1}^{N}\prod_{j=1}^{M} \left(\left(1 - \frac{b}{a}\right)e^{-aR_{i,j}} + \frac{b}{a}\right)^{\left(1-\frac{b}{a}\right)e^{-aR_{i,j}}+\frac{b}{a}} \qquad (4)$$

Now plug Equation (2) into Equation (4), we have:

$$L = \prod_{i=1}^{N}\prod_{j=1}^{M} \left(\left(1 - \frac{b}{a}\right)e^{-aU_i^T \cdot V_j} + \frac{b}{a}\right)^{\left(1-\frac{b}{a}\right)e^{-aU_i^T \cdot V_j}+\frac{b}{a}} \qquad (5)$$

Apply Stochastic Gradient Descent (SGD) to solve Equation (5) for a, b, U and V, and we acquired the following formulas :

$$\frac{\partial L}{\partial a} = \frac{bt_2t_5t_6}{t_7} - t_1t_4t_5t_6 - \frac{bt_5t_6}{t_7} - t_1t_{10}t_3t_8 + \frac{bt_{10}t_8}{t_7} - \frac{bt_8t_9}{t_7} \qquad (6)$$

, where :

$t_0 = \frac{b}{a}$, $t_1 = U_i^T \cdot V_j$, $t_2 = e^{-at_1}$, $t_3 = 1 - t_0$, $t_4 = t_2 \cdot t_3$, $t_5 = t_0 + t_4$, $t_6 = t_5^{t_0-1+t_4}$, $t_7 = a^2$, $t_8 = t_5^{t_5}$, $t_9 =$

$log(t_5), t_{10} = t_2 \cdot t_9$

, and :

$$\frac{\partial L}{\partial b} = \frac{t_3 \cdot t_4}{a} - \frac{t_1 \cdot t_3 \cdot t_4}{a} - \frac{t_1 \cdot t_5 \cdot t_6}{a} + \frac{t_5 \cdot t_6}{a} \quad (7)$$

, where :

$t_0 = \frac{b}{a}, t_1 = e^{-aU_i^T \cdot V_j}, t_2 = t_1(1 - t_0), t_3 = t_0 + t_2, t_4 = t_3^{t_0 - 1 + t_2}, t_5 = t_3^{t_3}, t_6 = log(t_3)$

, and :

$$\frac{\partial L}{\partial U_i} = -(at_3 t_4^{t_0 + t_3} V_j + at_1 t_2 t_4^{t_4} log(t_4) V_j) \quad (8)$$

, where :

$t_0 = \frac{b}{a}, t_1 = e^{-aU_i^T \cdot V_j}, t_2 = 1 - t_0, t_3 = t_1 \cdot t_2, t_4 = t_0 + t_3$

, and :

$$\frac{\partial L}{\partial V_j} = -(at_3 t_4^{t_0 + t_3} U_i + at_1 t_2 t_4^{t_4} log(t_4) U_i) \quad (9)$$

, where :

$t_0 = \frac{b}{a}, t_1 = e^{-aU_i^T \cdot V_j}, t_2 = 1 - t_0, t_3 = t_1 \cdot t_2, t_4 = t_0 + t_3$

To recover the predicted user item rating value, we use Equation (2) instead of Equation (1) because it is much simpler. We demonstrate the effectiveness of the GreyShot algorithm in the Experiment section. Based on the formulas derived for SGD, we notice that none of the equations contains input data information, therefore our algorithm is a Zeroshot Learning algorithm .

Unlike conventional school of zeroshot learning algorithms, which rely on transfer learning or meta learning to solve the problem, GreyShot has no reference to historic data or input data or data from other domain in the computation steps (Equation 6 - Equation 9) for its parameters. Therefore, it is an authentic zeroshot learning and privacy-preserving algorithm, whose effectiveness and competitiveness will be proved in the Experiment Section.

## 4. Evaluation metrics

Before we delve into the Experiment Section, we briefly overview the evaluation metrics in this section. We are about to use MAE (Mean Absolute Error) and Degree of Matthew Effect to evaluate the accuracy and fairness performance, respectively.

MAE (Mean Absolute Error) is defined as follows :

$$MAE = \frac{1}{M \times N} \sum_{i=1}^{M} \sum_{j=1}^{N} |R_{i,j} - \widehat{R}_{i,j}|$$

where M is the number of users and N is the number of items, $R_{i,j}$ is the rating value user i gives to item j, and $\widehat{R}_{i,j}$ is the predicted value of $R_{i,j}$ using recommendation algorithm.

Degree of Matthew Effect is defined in [39] as follows :

$$DME = 1 + n\left(\sum_{i=1}^{n} \ln \frac{x_i}{x_{max}}\right)^{-1}$$

Degree of Matthew Effect is a metric to capture the degree of skewness in the data distribution. The larger the value, the less skewed the data behaves. In the formula, $x_i$ represents the popularity rank of the i-th item, and $x_{max}$ is the maximum value of $x_i$.

## 5. Experiment

We tested our algorithm against DotMat, DotMat Hybrid, Classic Matrix Factorization, ZeroMat and Random Placement. We use 2 online public datasets for testing : MovieLens 1 Million Dataset [40] (6040 users and 3706 movies), and LDOS-CoMoDa Dataset [41] (121 users and 1232 movies). The accuracy metric we used is MAE (Mean Squared Error) , and the fairness metric we use is Degree of Matthew Effect [39].

Figure 1 shows that GreyShot is comparable with DotMat and DotMat Hybrid, and is slightly better than the classic matrix factorization algorithm. Figure 2 illustrates the effectiveness of Greyshot by fairness metric - it is comparable with the classic matrix factorization algorithm. From Figure 1 and Figure 2, we observe the effectiveness of the Greyshot algorithm.

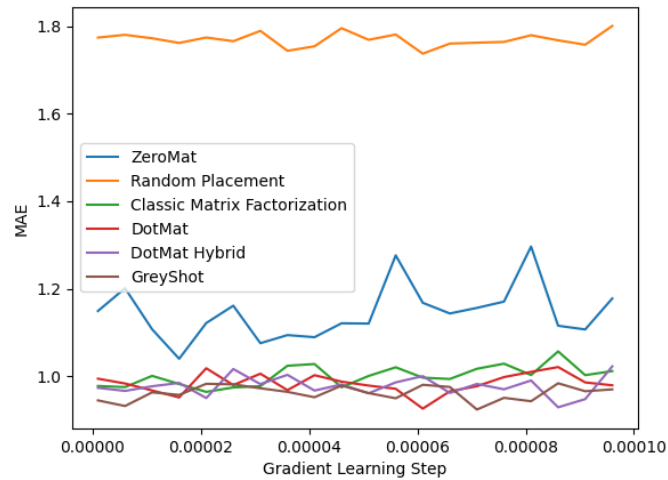

**Figure 1** Comparison of algorithms on MovieLens 1 Million Dataset : MAE Metric

Table 1 illustrates the minimum / average / maximum values of MAE scores on MovieLens 1 Million Dataset. It is obvious that GreyShot is the best on accuracy metric. Table 2 demonstrates that GreyShot is highly competitive on fairness metric.

Figure 3 and Figure 4 demonstrate the effectiveness of GreyShot. As in the previous experiment, GreyShot is a highly competitive algorithm among the different recommendation paradigms. It is safe to draw the conclusion that GreyShot is an effective and competitive Zeroshot and privacy-preserving technique.

**Table 1** Mininum/Average/Maximum MAE values of algorithms on MovieLens 1 Million Dataset

|         | Minimum       | Average       | Maximum       |
|---------|---------------|---------------|---------------|
| ZeroMat | -0.00649942   | -0.00635155   | -0.006217649  |
| Random  | -0.002516908  | -0.002516908  | -0.002516908  |
| MF      | -0.006562708  | -0.006334012  | -0.006059097  |
| DotMat  | -0.006508382  | -0.006334876  | -0.006177192  |
| DotMat H| -0.006581031  | -0.006351428  | -0.006137289  |
| GreyShot| -0.006535425  | -0.006367805  | -0.006121480  |

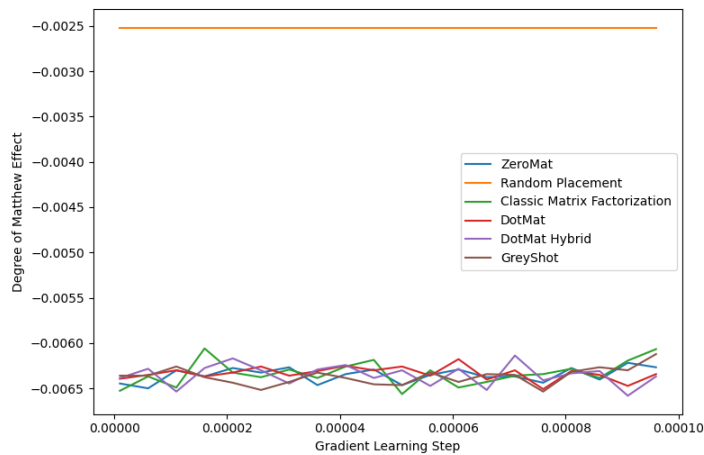

**Figure 2** Comparison of Algorithms on MovieLens 1 Million Dataset : Degree of Matthew Effect

**Table 2** Mininum/Average/Maximum Degree of Matthew Effect values of algorithms on MovieLens 1 Million Dataset

|         | Minimum      | Average      | Maximum      |
|---------|--------------|--------------|--------------|
| ZeroMat | 1.040034558  | 1.144713351  | 1.296571808  |
| Random  | 1.737323308  | 1.769572858  | 1.80067149   |
| MF      | 0.964436555  | 1.000656992  | 1.056944978  |
| DotMat  | 0.926127832  | 0.983751566  | 1.021156759  |
| DotMat H| 0.929148718  | 0.977833957  | 1.022756751  |
| GreyShot| 0.924078811  | 0.961816005  | 0.984081010  |

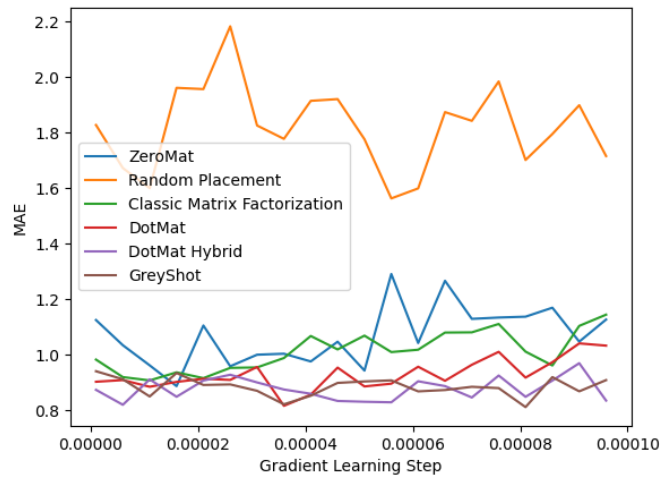

**Figure.3** Comparison of Algorithms on LDOS-CoMoDa Dataset : MAE Metric

**Table 3** Mininum/Average/Maximum MAE values of algorithms on LDOS-CoMoDa Dataset

|  | Minimum | Average | Maximum |
| --- | --- | --- | --- |
| ZeroMat | 0.88791712 | 1.070129181 | 1.29158364 |
| Random | 1.563622409 | 1.819836497 | 2.183021073 |
| MF | 0.908488139 | 1.012580896 | 1.145222931 |
| DotMat | 0.81712937 | 0.930345717 | 1.041829791 |
| DotMat H | 0.820879969 | 0.877978213 | 0.970522516 |
| GreyShot | 0.812457764 | 0.885428022 | 0.941657489 |

**Table 4** Mininum/Average/Maximum Degree of Matthew Effect values of algorithms on LDOS-CoMoDa Dataset

|  | Minimum | Average | Maximum |
| --- | --- | --- | --- |
| ZeroMat | -0.059993233 | -0.059031022 | -0.058980379 |
| Random | -0.058980379 | -0.058980379 | -0.058980379 |
| MF | -0.058980379 | -0.058980379 | -0.058980379 |
| DotMat | -0.058980379 | -0.058980379 | -0.058980379 |
| DotMat H | -0.058980379 | -0.058980379 | -0.058980379 |
| GreyShot | -0.058980379 | -0.058980379 | -0.058980379 |

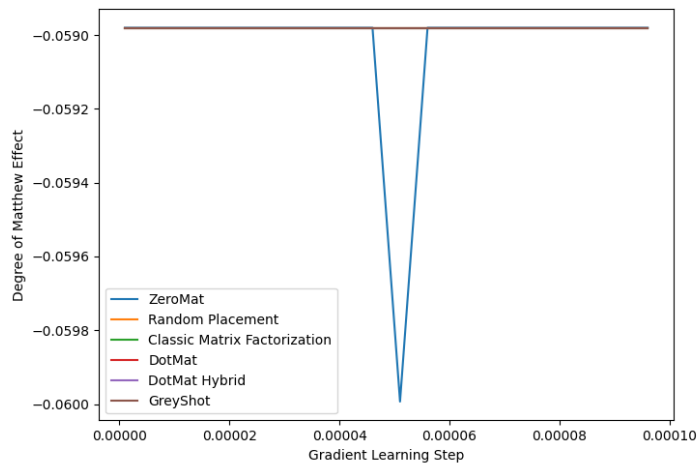

**Figure 4** Comparison of Algorithms on LDOS-CoMoDa Dataset : Degree of Matthew Effect

From Table 3 and Table 4, it can be observed that GreyShot is highly effective, when evaluated both by MAE and Degree of Matthew Effect metrics.

## 6. Discussion

Grey System Modeling is a technical field invented by Chinese scientist in 1982. Although the field is even older than recommender system itself, the number of researchers in the field is fewer than the number of scientists in recommendation. One of the reasons why we use Grey System Modeling to build recommender systems is to raise public awareness of the existence of such a research field. Another reason is that the experimental results are persuasive that Grey System Modeling is indeed a suitable technology for recommendation.

The theory of Grey System Modeling is deeply connected to first order differential equations. Some scientist believe that differential equations is the next hype after deep learning, and this time Grey System Modeling might become the next buzz word in both the academia and industry. First order differential equations seem to be expressive enough for a wide spectrum of problems in the real world, but differential equations of higher orders might also play a critical role in modeling other complex social and natural phenomena including recommender systems.

Our proposed algorithm GreyShot has been proved both theoretically and in experiments that it is an effective zeroshot learning and privacy-preserving algorithm that relies on no input data and is fairness-aware. In its computation step, it requires no data from the input, and when tested on the MovieLens and LDOS-CoMoDa dataset on both accuracy and fairness metrics, GreyShot exhibited highly competitive results.

## 7. Conclusion

Cold-start problem is a serious challenge for recommender system builders, and it remains an open problem even today. Although researchers have developed a series of authentic zeroshot learning algorithms after 2021, the number of research contributions in the field is still small compared with other paradigms such as transfer learning and meta learning. In this paper, we propose to use Grey System Modeling to solve the cold-start problem, and as a side-effect, the algorithm solves the privacy issue and therefore is an effective solution for the recommendation problem.

In future work, we would like to employ higher-order differential equations to solve the recommender system problem. We want to build competitive AI solutions with simple mathematics.